\documentstyle[here,epsfig]{mn}
\begin{document}
\def\simlt{\mathrel{\rlap{\lower 3pt\hbox{$\sim$}}\raise 2.0pt\hbox{$<$}}}
\def\simgt{\mathrel{\rlap{\lower 3pt\hbox{$\sim$}} \raise
2.0pt\hbox{$>$}}}

\title[] 
{Is there a Dichotomy in the Radio Loudness Distribution of Quasars ?}
\author[M. Cirasuolo,  A. Celotti, M. Magliocchetti, L. Danese] 
{M. Cirasuolo,  A. Celotti, M. Magliocchetti, L. Danese  \\ 
SISSA, Via Beirut 4, 34014, Trieste, Italy }
\maketitle \vspace {7cm }
\begin{abstract}
We present a new approach to tackle the issue of radio loudness in
quasars. We constrain a (simple) prescription for the intrinsic
distribution of radio-to-optical ratios by comparing properties of
Monte Carlo simulated samples with those of observed 
optically selected quasars. We find strong evidence for a dependence of the
radio luminosity on the optical one, even though with a large scatter.  The
dependence of the fraction of radio loud quasars on apparent and
absolute optical magnitudes results in a selection effect related to
the radio and optical limits of current surveys.\\ 
The intrinsic distribution of the radio-to-optical ratios shows a peak at 
$R^*_{1.4} \sim 0.3$,  with only $\simlt 5$~ per cent of objects being included
in a high $R^*_{1.4}$ tail which identifies the radio loud regime.
No lack or deficit of sources -- but only a steep transition region
-- is present between the radio loud and radio quiet populations at 
any $R^*_{1.4}$. 
We briefly discuss possible origins for this behaviour (e.g. absence of
jets in radio quiet sources, large range of radiative radio
efficiency, different life-times for the accretion and jet ejection
phenomena, ...).
\end{abstract}
%
\section{INTRODUCTION}
The origin of radio loudness of quasars is a long debated issue.
Radio observations of optically selected quasar samples showed only
10-40~\% of the objects to be powerful radio sources (Sramek \&
Weedman 1980; Condon et al. 1981; Marshall 1987; Miller, Peacock \&
Mead 1990; Kellermann et al. 1989). More interestingly, these early
studies suggested that quasars could be divided into the two
populations of ``Radio-Loud'' (RL) and ``Radio-Quiet'' (RQ) on the
basis of their radio emission.  Furthermore, Kellermann et al. (1989) found
that the radio-to-optical ratios, $R^*_{1.4}$ \footnote{Throughout
this paper, we will refer to the radio-to-optical ratio as that
between radio (1.4 GHz) and optical (B band) rest frame luminosities},
of these objects presented a bimodal distribution.  Miller, Peacock \&
Mead (1990) also found a dichotomy in the quasar population, although
this time radio luminosity was used as the parameter to define the
level of radio loudness.

In the last decade, our ability of collecting large samples of quasars
with faint radio fluxes has grown enormously, in particular thanks to
the FIRST (Faint Images of the Radio Sky at Twenty centimeters) Survey
at VLA (Becker, White \& Helfand 1995). However, despite the recent efforts,
radio loudness still remains an issue under debate. 
Works based on data from the FIRST survey (White et al. 2000;
Hewett et al. 2001) suggest that the found RL/RQ dichotomy could be
due to selection effects caused by the brighter radio and optical
limits of the previous studies.  On the contrary, Ivezic et al.
(2002) seem to find evidence for bimodality in a sample drawn from the Sloan
Digital Sky Survey. More recently Cirasuolo et al.  (2002)
(hereafter paper I) -- analyzing a new sample obtained by matching
together the FIRST and 2dF QSO Redshift Survey -- ruled
out the classical RL/RQ dichotomy in which the distributions of
radio-to-optical ratios and/or radio luminosities show a deficit of
sources, suggesting instead a smoother transition between the RL and
the RQ regimes.

Also from the interpretational point of view, the physical mechanism(s)
responsible for radio emission in Active Galactic Nuclei (AGN) is
still debated.  At least for RL quasars it is generally accepted to be
related to the process of accretion onto a central black hole (BH) --
the engine responsible for the optical-UV emission -- via the
formation of relativistic jets which can be directly imaged in nearby
objects.  The connection between the optical (accretion) and radio
(jet) emission is however unclear.  Phenomenologically there is
indication in RL objects of correlations between  radio emission
and luminosity in narrow emission lines, produced by gas
presumably photoionized by the nuclear optical--UV continuum (e.g.
Rawlings \& Saunders 1991). Results from paper I
show that, for a given optical luminosity, the scatter in radio power
is more than three orders of magnitude.  Note that if quasars
are AGN accreting near the Eddington limit, this result tends to
exclude the mass of the central BH as the chief quantity controlling
the level of radio activity (see also Woo \& Urry 2002), although we
cannot conclude anything on the possible presence of a threshold
effect, whereby RL AGN would be associated to the more massive BH
($M_{\rm BH} > 1-5 \times 10^9 M_{\odot}$, Laor 2000, Dunlop et al.
2002, but see also Woo \& Urry 2002 for a dissenting view).  On the
other hand, although controversial, there is some evidence for the
fraction of RL quasars to increase with increasing optical luminosity
(Padovani 1993; La Franca et al. 1994; Hooper et al. 1995; Goldschmidt
et al. 1999; paper I; but see also Ivezic et al. 2002 for a different
view).

Clearly, the uncertainties on the presence of a dichotomy, the
character of radio loudness and the consequent poor knowledge of its
origin (dependence on BH mass, optical luminosity etc.) are due to the
analysis of different samples, often very inhomogeneous because of
selection effects both in the optical and radio bands, i.e. the lack
of a single sample covering all the ranges of optical and radio
properties of quasars.

Therefore, in order to shed light on this issue, we adopted the
alternative approach of starting from simple assumptions on the
intrinsic properties of the quasar population as a whole -- namely an
optical quasar luminosity function and a prescription to associate a
radio power to each object - and, through Monte Carlo simulations,
generate unbiased quasar samples. By applying observational limits in
redshift, apparent magnitude and radio flux we can then compare the
results of the simulations with the properties of observed samples.
The aim of this approach is of course twofold: constrain the initial
hypothesis on the intrinsic nature of quasars, by requiring
properties of the simulated samples -- such as $R^*_{1.4}$ and radio
power distributions, fraction of radio detections etc. -- to be in
agreement with the observed ones; test the effects of the
observational biases on each sample by simply changing the
observational limits.

The layout of the paper is the following. In Section 2 we briefly
present the samples used to constrain the models, while in Section 3
we describe the procedures adopted to generate simulated samples.  In
Section 4 results of the simulations and comparisons of the
properties of the observed and simulated samples are shown.  We
discuss the physical meaning of these results and summarize our
conclusions in Section~5.  Throughout this work we will adopt $H_0 =$
50 km s$^{-1}$ Mpc$^{-1}$, $q_0 = 0.5$ and $\Lambda = 0 $.
\section{The Datasets}
As there is not (yet) a single sample able to cover, with enough
statistics and completeness, the total range of known radio activity
(e.g. the distribution of radio-to-optical ratios and/or radio powers), we
have to consider various samples to constrain - as completely as
possible - the properties of our simulated samples.  We choose three
samples of optically selected quasars, namely the 2dF Quasar Redshift
Survey (2dF), the Large Bright Quasar Survey (LBQS) and the Palomar Bright 
Quasar Survey (PBQS).
These have in fact a high completeness level and are quite
homogeneous since all of them are optically selected in the blue
band and have a similar radio cut.  The 2dF and LBQS have been
cross-correlated with the FIRST survey with a limiting flux at 1.4 GHz of
1~mJy, while the PBQS has been observed at 5 GHz down to a
limiting flux of 0.25~mJy, which is comparable to the 1~mJy flux limit
at 1.4 GHz, for a typical radio spectral index $\alpha_{\rm R} = 0.8$
\footnote{Throughout this work we define the radio spectral index as
$S_{\nu} \propto \nu^{-\alpha_{\rm R}}.$}. Equally crucial for our
work is the fact that these samples produce a very large coverage of
the optical luminosity-redshift plane and provide information on
different regimes of radio activity (see Fig.\ref{distr_R}). \\

Here we briefly describe the main characteristics of the three
samples. In order to have complete homogeneity and favour the
comparison with simulations, for each sample we have only considered
the ranges in redshift, apparent magnitude and radio flux with the
highest completeness.  We have also selected sources with $M_{\rm B}
\leq -23$ to avoid contaminations from the host galaxy light (Croom et
al. 2001; paper I) and we have applied a mean correction of 0.07 mag
to convert $b_{\rm J}$ to B magnitudes (see paper I), on the basis of
the composite quasar spectrum compiled by Brotherton et al. (2001).
The same composite spectrum has also been used to compute the
k-correction in the B band.
%
\subsection{2dF Quasar Redshift Survey}
We have used the first public release of the 2dF QSO Redshift Survey, the so
called {\it 2QZ 10k catalogue}.  Here we briefly recall its main
properties, while a complete description can be found in Croom et al.
(2001). QSO candidates with $18.25 \le b_{\rm J} \le 20.85$ were
selected from the APM catalogue (Irwin, McMahon \& Maddox 1994) in two
$75^{\circ} \times 5^{\circ}$ declination strips centered on
$\delta=-30^\circ$ and $\delta=0^\circ$, with colour selection
criteria $(u-b_{\rm j})\le 0.36$; $(u-b_{\rm j})< 0.12-0.8\;(b_{\rm
j}-r)$; $(b_{\rm j}-r)< 0.05$.  Such a selection guarantees a large
photometric completeness ($ > 90 $ per cent) for quasars within the
redshift range $ 0.3 \le z \le 2.2$.  The final catalogue contains
$\sim 21,000$ objects with reliable spectral and redshift
determinations, out of which $\sim 11,000$ are quasars ($\sim 53$~per
cent of the sample).

As extensively described in paper I, this sample has been
cross-correlated with the FIRST survey (Becker et al. 1995).  The
overlapping region between the FIRST and 2dF Quasar Redshift Surveys
is confined to the equatorial plane: $ 9^h \; 50^m \leq {\rm RA(2000)}
\leq 14^h \; 50^m$ and $ -2.8^{\circ} \leq {\rm dec(2000)} \leq
2.2^{\circ}$.  For the matching procedure a searching
radius of 5 arcsec and an algorithm to collapse multiple-component
radio sources (jets and/or hot-spots) into single objects were used.
(Magliocchetti et al.  1998).  The resulting sample is constituted by
113 objects, with optical magnitudes $18.25 \le b_{\rm j} \le 20.85$
and radio fluxes at 1.4 GHz $S_{1.4 {\rm GHz}}\ge 1$~mJy, over an
effective area of 122.4 square degrees. In the following we consider
the sub-sample (hereafter called the FIRST-2dF sample) containing
89 objects with absolute magnitudes brighter than $M_{\rm B} \leq -23$ spanning
in the redshift range $ 0.35 \leq z \leq 2.1$.
%
\subsection{Large Bright Quasar Survey}
The Large Bright Quasar Survey (LBQS) comes as the natural extension
to brighter magnitudes of the 2dF sample since it has been derived with
the same selection criteria.  A detailed description of this survey
can be found in Hewett et al.  (1995). It consists of quasars
optically selected from the APM catalogue (Irwin et al. 1994) at
bright ($b_{\rm j} < 19$) apparent magnitudes. Redshift measurements
were subsequently derived for 1055 of them over an effective area of
483.8 square degrees. Due to the selection criteria of the survey,
quasars were detected over a wide redshift range ($0.2 \le z \le
3.4$), with a degree of completeness estimated to be at the $\sim 90$
per cent level.\\ 
More recently, this sample was cross-correlated with
the FIRST survey (Hewett et al. 2001) by using a searching radius of
2.1 arcsec over an area of the sky of 270 square degrees. This
procedure yielded a total of 77 quasars with radio fluxes $S_{1.4 {\rm
GHz}} \ge 1$~mJy, magnitudes in the range $16 \simlt b_{\rm j} \simlt
19$ and with a fractional incompleteness of $\sim 10$ per cent.  For
homogeneity with the FIRST-2dF, out of this sample we have only considered 58
sources with apparent magnitudes $16 \leq b_{\rm j} \leq 18.8 $ and
redshifts $ 0.2 \leq z \leq 2.2$, all of them with $M_{\rm B}
\leq -23$ (hereafter called the FIRST-LBQS sample).
%
\subsection{Palomar Bright Quasar Survey}
%
We have also considered the Palomar Bright Quasar Survey (Schmidt \& Green
1983), which represents one of the historical and most studied sample
of quasars, in order to cover the very bright end of the absolute
magnitude distribution.  It contains 114 objects brighter than an effective
limiting magnitude $B = 16.16$, selected  over an area of 10,714 square
degrees using the UV excess technique. Such quasars have been
observed using the VLA at 5 GHz with 18 arcsec resolution, down to a
flux limit of 0.25 mJy ($4\sigma$) (Kellermann et al. 1989).  In order
to reduce, as much as possible, the contamination due to the host
galaxy, we have chosen not to consider the very local sources with
redshift $z < 0.1$.  Our choice is also justified by the fact that all
these objects are classified as ``galaxy'' in the NED archive
(NASA/Ipac Extragalactic Database), indicating that the host galaxy is
clearly visible.  Again, for homogeneity with the former datatsets, 
from this sample we have only selected objects with $13
\leq B \leq 16.16$, in the redshift range $ 0.1 \leq z \leq 1.5$, with
$M_{\rm B} \leq -23$ and radio flux $S_{\rm 5 GHz} \geq 0.25$ mJy,
ending up with a sub-sample of 48 radio sources (hereafter called the
PG sample).
\begin{figure}
\center{{
\epsfig{figure=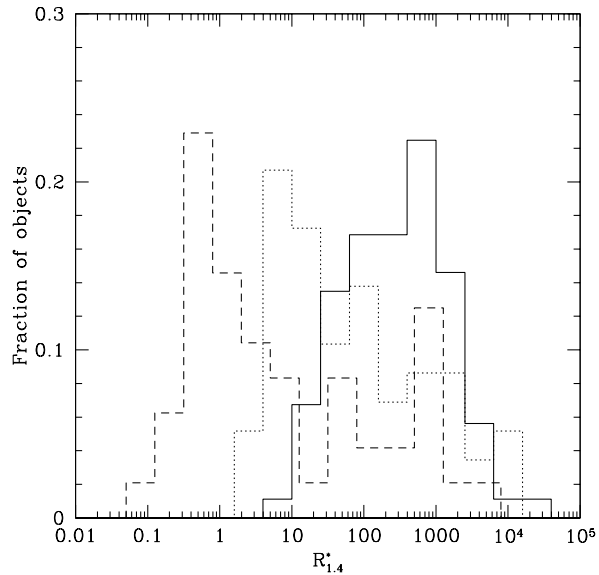,height=8cm}
}}
\caption{\label{distr_R} Distributions of the radio-to-optical ratios
for the three samples: FIRST-2dF (solid line), FIRST-LBQS (dotted
line) and PG (dashed line)}
\end{figure}
%
\section{Monte Carlo Simulations}
%
As already mentioned, as selection effects both in the optical and
in the radio bands bias each of the samples introduced in Section 2
with respect to the quasar
radio properties, our approach is to start from simple assumptions on
the intrinsic features of the quasar population and test them by comparing 
results from simulated samples -- generated through Monte
Carlo realizations -- with those of the observed samples.  We decided
to assume as the two fundamental ingredients to describe the 
quasar population a well defined Optical Luminosity Function (OLF)
- from which to obtain redshift and optical magnitude for the
sources - and a distribution of radio-to-optical ratios which provides each
source with a radio luminosity.\\
Here we briefly specify these assumptions and
describe the procedure applied in our Monte Carlo simulations.
%
\subsection{Optical Luminosity Function}\label{olf}
The best determined OLF for the whole quasar population currently
available is the one obtained from the 2dF Quasar Redshift Survey
(Boyle et al. 2000; Croom et al. 2001), based on $\sim 11,000$
sources. This can be described as a broken power-law (Croom et al.
2001)
\begin{equation}
\phi(M_{\rm B},z) = \frac{\phi^*}{10^{0.4 [C(z)(\alpha +1) ]}  + 
10^{0.4 [C(z)(\beta +1) ]}}, 
\end{equation}
with $\alpha= 3.28$, $\beta=1.08$, $C(z) = M_{\rm B} - M_{\rm B}^*(z)$,
where sources undergo a pure luminosity evolution parametrized by the
expression
\begin{equation}
M_{\rm B}^*(z) = M_{\rm B}^*(0) - 2.5 (k_1 z + k_2 z^2),
\end{equation}
with $k_1=1.41$, $k_2=-0.29$ and $ M_{\rm B}^*(0) = -21.45$.
One of the main advantages of using this
luminosity function is that it has been obtained in the B band and
over the wide redshift range $0.35 \leq z \leq 2.2$, therefore reducing
the need for extrapolations, which would add further uncertainties.
Some extrapolation is instead needed for the bright end of the 
luminosity function ($M_B \simlt -27$), which is not well sampled by the
available data.
Adopting the above OLF
we have then randomly generated sources, over redshift and apparent
magnitude ranges necessary to compare them with each of the three
samples.

Note that, we have used the Optical LF as a description of the key quasar
properties and number density mainly because we are going to compare
our results with samples of optically selected quasars. 
In a more general picture
one should also include the contribution to the quasar population
given by the obscured (type II) QSO. To this aim it would be
plausibly more appropriate to use a hard X-ray LF, which however so
far has only been computed for type I objects (La Franca et al.  2002)
and which thus results to be similar in shape and evolution to the one
determined in the soft X-ray band (Miyaji et al 2000).  We stress that,
 since the radio emission is not affected by obscuration, the contribution of
type II quasars would affect our results (e.g. the distribution of
radio-to-optical ratios) only if there was a dependence of the
obscuration level on the quasar optical luminosity. However, as recent
results (La Franca et al. 2002; Tozzi et al. 2001; Giacconi et al.
2002; Rosati et al. 2002) seem to suggest, the fraction of these
obscured quasars is expected to be small when compared to the unobscured ones.
The above conclusion implies that the presence of type II quasars should not 
significantly affect our findings and therefore we have decided to neglect 
their (uncertain) contribution.
%
%
\subsection{Radio vs Optical Luminosity}
\begin{figure}
\center{{
\epsfig{figure=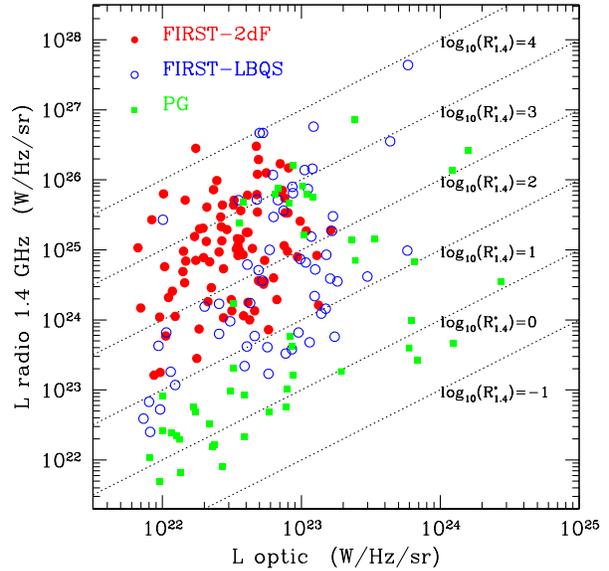,height=8cm}
}}
\caption{\label{lolr} Optical versus radio luminosity for sources in
the considered samples. The dotted lines are loci of constant
radio-to-optical ratio.}
\end{figure}

The other ingredient needed for our analysis is a relation between optical 
and radio luminosities. As already concluded in paper I, although present,
this shows a wide spread:
Figure \ref{lolr} illustrates -- for the objects in the three samples
considered in this work -- how sources with a particular optical
luminosity can be endowed with radio powers spanning up to three
orders of magnitude.

We have thus taken into account two (simple) different scenarios. The first
one assumes a relation between radio and optical luminosity,
although with a large scatter (Model A); in this case we use a
distribution in $\log_{10}(R^*_{1.4})$ and compute the radio power as
$\log_{10}(R^*_{1.4}) + \log_{10}(L_{\rm opt})$. The second case
describes the possibility for the radio luminosity to be completely
unrelated to the optical one, and assume a distribution in
$\log_{10}(L_{\rm radio})$ for any given optical luminosity (Model B). For
Models A and B respectively, we then consider different shapes for the 
distribution of radio-to-optical ratios and radio luminosities:
 a) the simplest case of a flat uniform distribution
over the whole range of radio-to-optical ratios or radio powers; b) a
single Gaussian distribution, in which the RL regime could simply
represent its tail; c) two Gaussians, the first peaked in the RL
regime and the second in the RQ one, in order to allow for a more
flexible shape of the distributions and test the hypothesis of
bimodality.

As a final step, radio flux densities have been computed from radio
powers by assuming 80 per cent of these sources to be steep
spectrum ($\alpha_{\rm R} = 0.8$), while the others to be flat
spectrum ($\alpha_{\rm R} = 0$) objects.  These fractions have been
chosen according to the results found in paper I: although for the
FIRST-2dF sample we only managed to compute the radio spectral index
for the ten most luminous objects, roughly finding the same fraction
of steep and flat spectra, the remaining sources do not have any
counterpart at radio frequencies $> 1.4$ GHz, suggesting that most of
the quasars have a steep spectrum.  These assumed fractions are also
in agreement with that of steep spectrum sources observed in the PG
sample.
It must be noticed that in these associations of a radio luminosity to
each optical quasar there is a strong implicit assumption on the
evolution, namely that radio and optical luminosities evolve in the
same way. Even though we have no direct evidence, some hints
supporting this hypothesis can be bound from the data. On one side in paper I it
has been shown that the $R^*_{1.4}$ distribution, both for sources in
the FIRST-2dF and in the FIRST-LBQS, is completely independent of redshift.
This is not expected if radio and optical luminosities had
significantly different evolutions.  On the other side, an independent
indication of such hypothesis is obtained from the $<V/V_{\rm max}>$
test, which we applied separately to radio detected objects (from the
FIRST-2dF) and to quasars representative of the population as a whole
(from the total 2dF).  For objects in the radio sample we computed
$V_{\rm max}$ by using the maximum redshift at which a source could have
been included both in the radio and in the optical datasets, given the
radio flux and optical magnitude limits. We considered three redshift
bins and the results \footnote{The errors on the mean value of
$<V/V_{\rm max}>$ have been computed as $(\sqrt{12 \: N})^{-1}$, where
N is the number of objects in each bin.} are reported in Table
\ref{tab:vmax}.  The values of $<V/V_{\rm max}>$ for the two
populations are perfectly compatible in all redshift bins, even though
the errors for the radio sample are larger due to the smaller number
of objects.  This again suggests the evolutionary behaviour of the
radio detected and total populations to be similar, justifying our
assumption.
\begin{table}
\begin{center}
\begin{tabular}{lcc} \hline \hline
 z              &  Total                & Radio \\ \hline
 0.35 - 1       & $0.61 \pm 0.01$       & $0.58 \pm 0.07$ \\
 1 - 1.5        & $0.55 \pm 0.01$       & $0.53 \pm 0.05$ \\
 1.5 - 2.1      & $0.51 \pm 0.008$      & $0.50 \pm 0.05$ \\
\hline \hline
\end{tabular}
\end{center}
\caption{\label{tab:vmax} Mean $V/V_{\rm max}$ values for the
radio detected and the total quasar population from the 2dF sample as measured
in three different redshift bins.}
\end{table}
%
\subsection{Constraints from the data}\label{constrain_data}
\begin{figure}
\center{{
\epsfig{figure=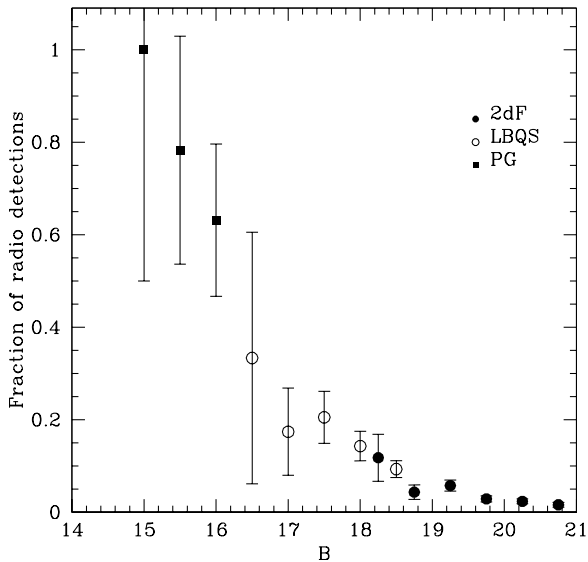,height=8cm}
}}
\caption{\label{frac_B} Fraction of quasars with a radio counterpart
in the three samples as a function of the apparent optical magnitude.}
\end{figure}

Adopting each of the six models described in the previous section and
applying the observational limits (reported for clarity in Table
\ref{tab:data}), we have simulated samples of radio emitting quasars.
Comparisons with the FIRST-2dF, FIRST-LBQS and PG samples provide
several constraints. In particular the simulated populations must
reproduce the observed:
\begin{itemize}
\item distributions of radio-to-optical ratios and radio powers.
The three samples identify different levels of radio activity: while the
FIRST-2dF -- which is the optically faintest -- well traces the RL
regime for $R^*_{1.4} \simgt 50$, the FIRST-LBQS is sensible to the RL-RQ
transition and the PG is particularly suited
to constrain the RQ regime (see Figure \ref{distr_R}).
\item fraction of radio detections. As shown in Figure \ref{frac_B},
this fraction depends on the optical limiting magnitude of the survey:
from $\sim 3$ per cent for the optically faint sources (FIRST-2dF), to
$\sim 20$ per cent for objects with $B \sim 17$ (FIRST-LBQS) and up to
70 per cent or more at the brightest magnitudes (PG). \\ 
Note that a similar dependence is also followed by the intrinsic luminosity,
as the fraction of radio detections grows from $\simlt 3$ per cent at $M_{\rm
B} \sim -24$ up to 20-30~per cent for the brightest $M_{\rm B} \sim
-28$ objects (paper I; Padovani 1993; La Franca et al. 1994; Hooper et
al. 1995; Goldschmidt et al. 1999).
\item number counts, both in the radio and in the optical band.
\item redshift and absolute magnitude distributions.
\end{itemize}
\begin{table*}
\begin{center}
\begin{tabular}{llcccc} \hline \hline
Survey  & $N_{\rm QSO}$ & $z_{\rm lim}$ & $B_{\rm lim}$ & $S_{\rm lim}$ (mJy) &Area (deg$^2$) \\ 
\hline
FIRST-2dF     & 89 & $0.35 \leq z \leq 2.1$& $18.25 \leq b_{\rm j} \leq 20.85$ &$S_{\rm 1.4\: GHz} \geq 1 $ & 122.4\\
FIRST-LBQS    & 58 & $0.20 \leq z \leq 2.2$& $16.00 \leq b_{\rm j} \leq 18.80$ &$S_{\rm 1.4\: GHz} \geq 1 $  & 270.0\\
PG            & 48 & $0.10 \leq z \leq 1.5$& $13.00 \leq B \leq 16.16$ &$S_{\rm 5 \: GHz} \geq 0.25 $  & 10714.0\\
\hline \hline   
\end{tabular}
\end{center}
\caption{\label{tab:data} Selection limits and covered areas for the
surveys used in our analysis. $N_{\rm QSO}$ is the number of quasars
with $M_{\rm B} \leq -23$}
\end{table*}
%
\section{Results}
Monte Carlo simulations have been run exploring the widest range
of values for the free parameters -- which describe the distributions
of radio-to-optical ratios (Model A) or radio powers (Model B) -- in
order to find all the sets of values able to reproduce the data.  
We have used simulated samples with a hundred times more objects than the original 
datasets and the realizations have been repeated with different initial seeds, 
in order to minimize the errors on the simulated quantities.
We tested the validity of each model by comparing the properties of
simulated and observed samples through statistical tests: a
Kolmogorov-Smirnov (KS) test for the $R^*_{1.4}$ and radio power
distributions, and a $\chi^2$ test for the fraction of radio
detections as a function of apparent and absolute magnitude, and also
for the optical and radio number counts.
%
\subsection{Results for flat and single Gaussian distributions}
We find that the simplest model, a flat distribution, is totally
inconsistent with the data and rejected by the statistical tests, both
for Model A and B. In particular it is unable to reproduce the
observed number counts for the three samples simultaneously: the data,
in fact, require the number of RL quasars to be less than the RQ ones,
while this distribution spreads objects uniformly.\\
We then tested
the single Gaussian distribution, which also fails, for both Model A
and B, mainly because it is unable to simultaneously reproduce the observed
distribution of $R^*_{1.4}$ in the RL regime and the total number of objects in 
the three  samples.
%
\subsection{Results for the two-Gaussian distribution}\label{twogauss}
\begin{figure*}
\center{{
\epsfig{figure=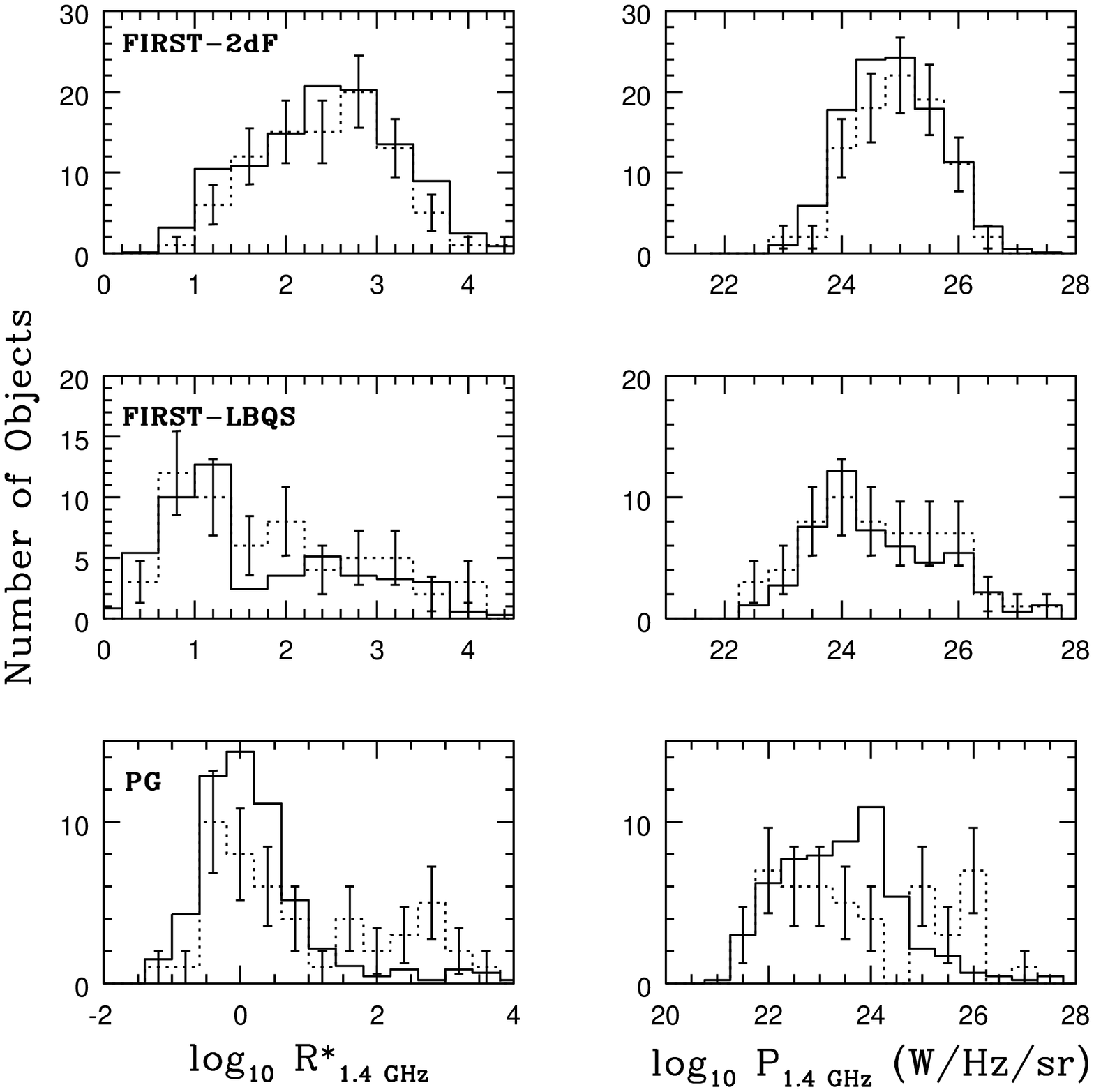,height=12cm}
}}
\caption{\label{model1} Comparison between the observed (dotted lines)
and the predicted (solid lines) distributions of radio-to-optical
ratios (on the left) and radio powers (on the right) for the three
samples.  Error-bars on the observed distributions are derived by assuming a 
Poissonian distribution.}
\end{figure*}

\begin{figure*}
\center{{
\epsfig{figure=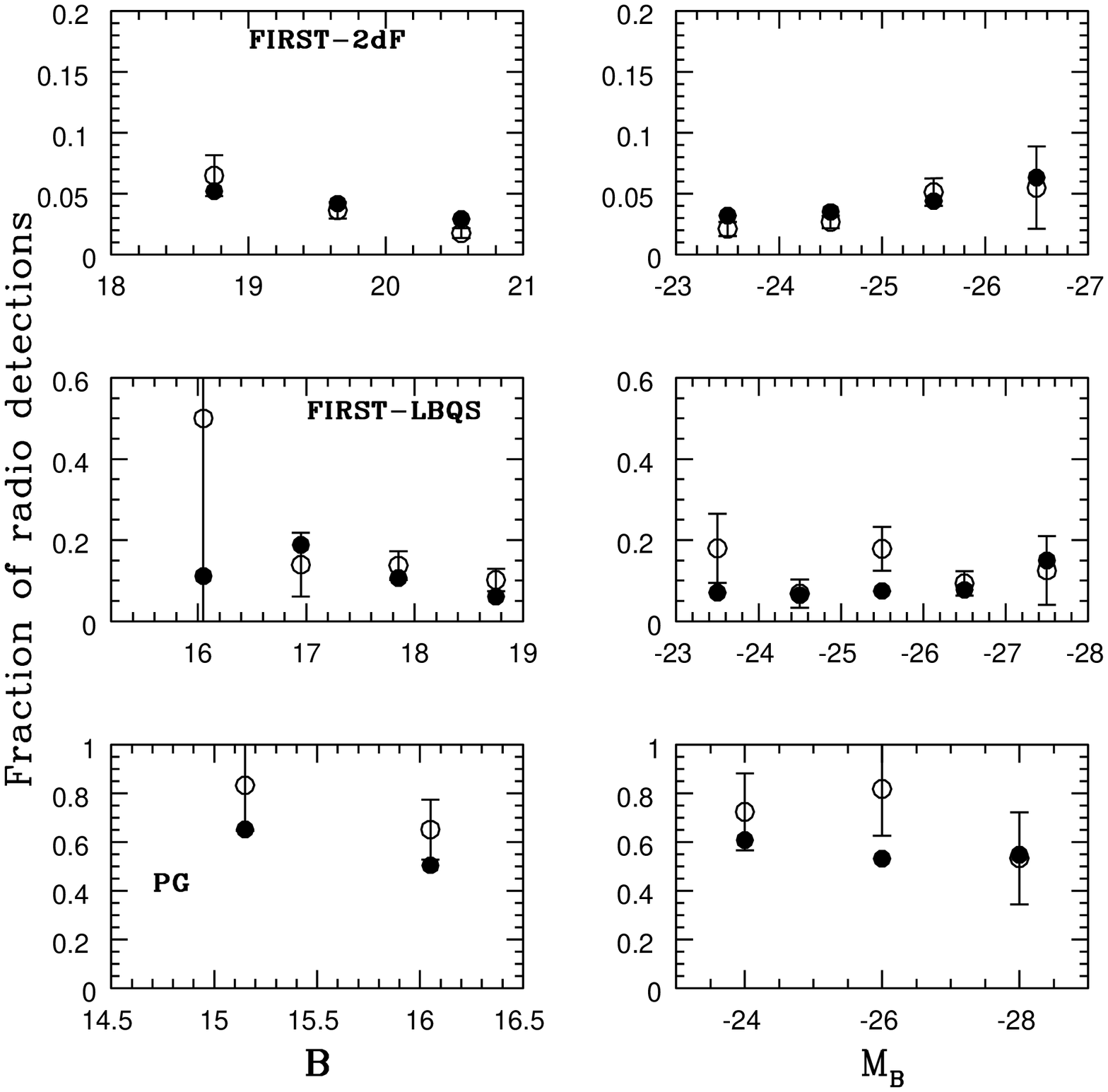,height=8cm}
\epsfig{figure=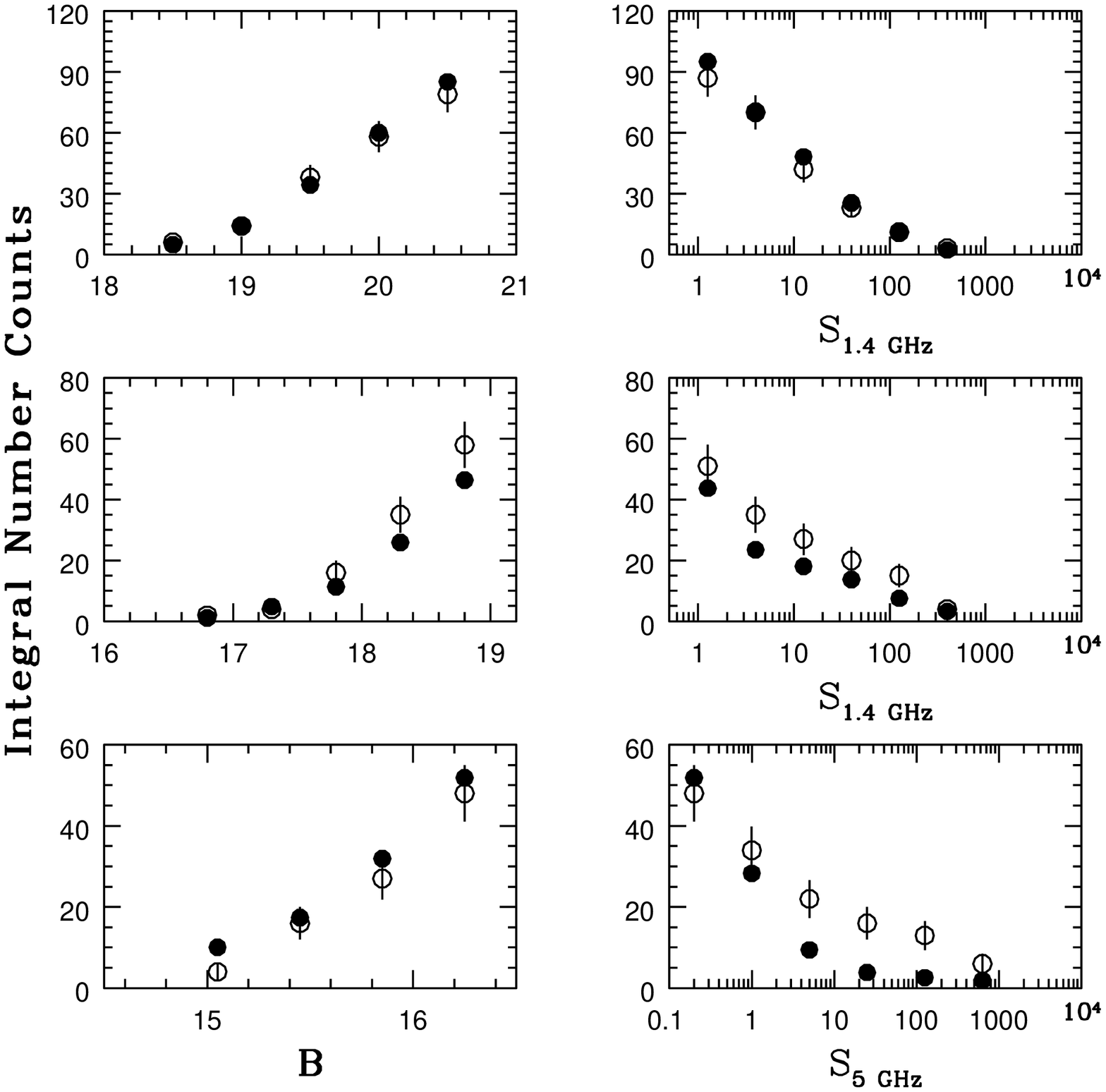,height=8cm}
}}
\caption{\label{model2} Left panels: Comparison between the observed
(open circles) and the predicted (solid dots) fraction of radio
detections for the three samples, as a function of the apparent and
absolute magnitudes.  Right panels: Comparison of the optical and radio
integral number counts. Simbols are as in the former case.}
\end{figure*}

\begin{figure*}
\center{{
\epsfig{figure=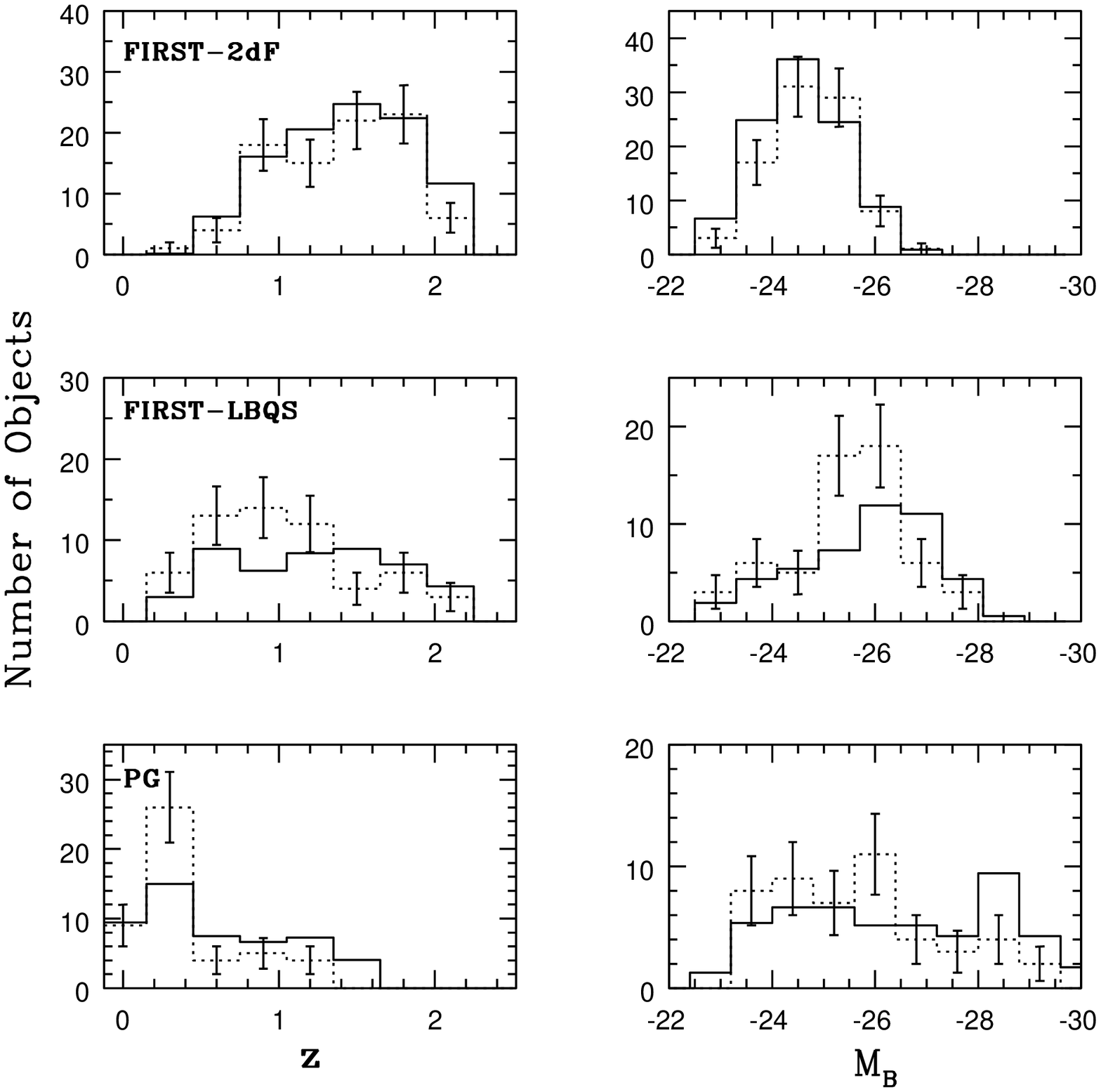,height=12cm}
}}
\caption{\label{distr_M_z} Comparison between the observed (dotted
lines) and expected (solid lines) distributions of redshift (on the
left) and absolute magnitude (on the right). }
\end{figure*}

Given the above difficulties, we considered the more flexible model of
two Gaussians, respectively centered in the RL and RQ regions. As said, each 
one of the considered samples traces different radio regimes: in particular
the FIRST-2dF sample well constrains the shape of the distribution at
high $R^*_{1.4}$, and also -- through the number counts -- well
determines the fraction of objects with high levels of radio emission.
As a consequence, the parameters of the first Gaussian (center $x_1$
and dispersion $\sigma_1$) and the relative fraction of objects in the
two Gaussians are mainly determined by this sample. The other two
samples constrain parameters of the second Gaussian in the RQ regime:
the PG sample mainly determines its peak position; the FIRST-LBQS sample --
which traces the transition region between RL and RQ -- constrains
quite well the shape of the wing of the second Gaussian and its
overlap with the first one.
\begin{table}
\begin{center}
\begin{tabular}{ccccc} \hline \hline
 $x_1$          & $\sigma_1$    & $x_2$ & $\sigma_2$ & Fraction\\ \hline
 $ 2.7 \pm 0.2$ & $0.7 \pm 0.2$ & $-0.5 \pm 0.3$ & $0.75 \pm 0.3$ &$97 \pm 2$~per cent\\
\hline \hline
\end{tabular}
\end{center}
\caption{\label{tab:mc} Best fit parameters for Model A, expressed in
$\log_{10}\: R^*_{1.4}$. $x_1$ and $\sigma_1$ are the center and
dispersion of the Gaussian in the RL regime, while $x_2$ and
$\sigma_2$ are those for the Gaussian in the RQ one. ``Fraction''
indicates the percentage of objects having radio-to-optical ratios described by 
the second Gaussian. Errors have been obtained as explained in 
Section \ref{twogauss}}
\end{table}
\begin{figure}
\center{{
\epsfig{figure=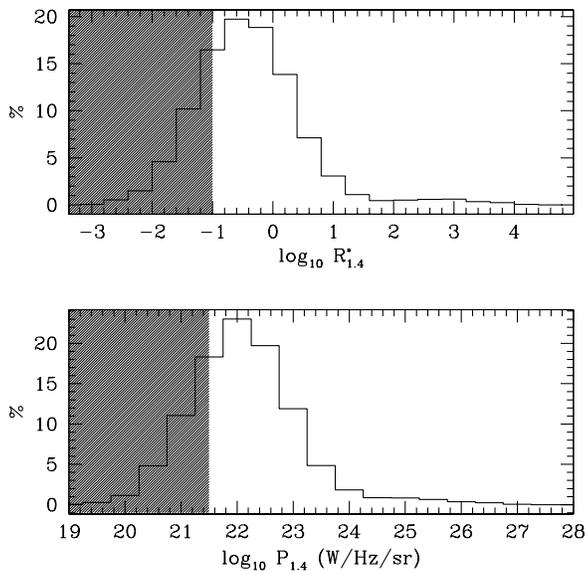,height=8cm}
}}
\caption{\label{shape_R_P} Distribution of radio-to-optical ratios
(top panel) and of radio powers (bottom panel) obtained from the best-fit 
set of parameters of Model A (see Table \ref{tab:mc}). 
The distributions are plotted in
binned form and the shaded regions indicate the range of $R^*_{1.4}$
and $P_{1.4}$ for which no data are available.}
\end{figure}
\begin{itemize}
\item Model B. In this case, i.e. radio luminosity completely
unrelated to the optical one, no set of parameters has been found able
to satisfy the statistical tests for all the observed quantities. 
In particular, this model is not able to
predict the growing fraction of radio detections as a function of
both apparent and absolute magnitudes, and to reproduce in a satisfactory way
the shapes of the radio-to-optical ratio distributions simultaneously
in the case of  FIRST-2dF and FIRST-LBQS.\\
  
\item Model A. A promising solution has been found in this case,
namely by assuming radio and optical luminosities to be related even
though with a large scatter. The radio-to-optical ratio and radio
power distributions corresponding to this solution are displayed in
Figure \ref{shape_R_P} and the model parameters are given in Table
\ref{tab:mc}.  A comparison between the properties of observed and
simulated samples is shown in Figures \ref{model1} and \ref{model2}.
We find a good agreement with the $R^*_{1.4}$ and radio power
distributions of the FIRST-2dF and FIRST-LBQS samples (high KS
probabilities, from 0.2 up to 0.9, for the distributions in Figure
\ref{model1}). The simulated datatset is also able to reproduce the observed
fraction of radio detections, both as a function
of apparent and absolute magnitudes (with a significance level for the
$\chi^2$ test $ > 0.05$ see Figure \ref{model2}) and the number counts,
except for a tendency -- compatible within the errors -- to
over-estimate the FIRST-2dF and correspondingly under-estimate the
FIRST-LBQS counts.\\
However, substantial disagreement is found for
the $R^*_{1.4}$ and radio power distributions when compared to the observed
quantities in the PG sample (KS
probabilities $< 10^{-2}$).  The discrepancy is mainly due to the
presence in this sample of $\sim 10$ objects with high values of
$R^*_{1.4}$ ($ \simgt 100$), which also determine the different shape
of the radio number counts (see Figure \ref{model2}). We stress that
we found no solution compatible with both the PG and the other two
samples: the excess of RL sources in the PG dataset is in fact inconsistent
with the both FIRST-2dF and FIRST-LBQS samples.  We therefore investigated
in more details the properties of these $\sim 10$ PG sources, by
looking at their radio and optical-to-X-ray spectral indices and the
compactness of the radio emission. We found the properties of these
quasars to be completely indistinguishable from those of other
objects in the PG sample, except for a slightly different redshift
distribution (these RL quasars are mainly concentrated in the range
$0.3 \leq z \leq 0.5$) and (by definition) their high $R^*_{1.4}$. We
notice however that it has been shown that the PG sample is incomplete
(Goldschmidt et al.  1992) and it has also been suggested this
incompleteness not to be random with respect to the radio properties
(Miller, Rawlings \& Saunders 1993; Goldschmidt et al.  1999). Because
of this we are more confident that the FIRST-2dF and FIRST-LBQS are better
suited to represent the shape of the $R^*_{1.4}$ distribution, at least in the
RL regime.
Due to these problems with the PG sample, we have therefore not considered
the constraints on the $R^*_{1.4}$ and radio power distributions for this
sample, converging to the same best fit model as discussed before. 

As a further test we have looked at the redshift and absolute magnitude
distributions. The comparison between data and model predictions is shown in 
Figure \ref{distr_M_z}, which reveals an excellent agreement 
in the case of the FIRST-2dF sample. A worse accordance has
been found for the other datasets, even though the simulated vs. observed 
distributions are  still compatible within the errors.
The simulated samples (in these two cases) reveal a tendency to overproduce 
objects at bright ($M_B \simlt -27$) absolute magnitudes, 
probably due to the extrapolation of the bright end of 
the luminosity function (see Section \ref{olf}). However, we want to stress that
this effect cannot explain the disagreement in the $R^*_{1.4}$ and radio power
distributions found for the PG sample. 
In fact, those RL sources found to be in excess with respect to the expectations
-- and clearly visible 
in the redshift bin at $z \sim 0.4$ -- have absolute magnitudes 
$-25 \simlt M_B \simlt -26$, a range in which the OLF is well defined. 
\end{itemize}

Note that the above results have been obtained by relaxing the constraints for
the PG sample only on the $R^*_{1.4}$ and radio power distributions,
while still considering in the analysis the observed total number of objects 
and the fraction of radio detections: in fact, by completely  eliminating
this sample we would loose important and robust constraints on the RQ
regime.

Finally we evaluate the ranges of acceptable parameters by varying
their best fit values until one of the constraints (e.g.  $R^*_{1.4}$
and radio power distributions, fraction of radio detections, etc.) was
no longer satisfied with respect to the KS and/or $\chi^2$ tests. We
adopted, as acceptable limits, a probability of 0.1 for the KS test
and a confidence level of 0.05 for the $\chi^2$ test. In this way we
obtained a reference error estimate on the parameters which is given
in Table \ref{tab:mc}.
%
\section{Discussion}
%
The first point worth stressing is the ``uniqueness'' of the
solution found. The combination of all the observational constraints
is very cogent and thus, despite large errors on each constraint, we
find that only one solution in the whole is able to simultaneously reproduce
measurements from the three surveys. Furthermore, 
the uncertainties associated to the various parameters are in this case 
relatively small (see Table \ref{tab:mc}).  \\ 
It is also intriguing to notice that a simple
prescription for the distribution of the radio-to-optical ratios is
able to well reproduce all of the available observational constraints
given by three different samples.  It is important to remark here that
in order to reproduce the
data we need a dependence of the radio luminosity on the optical one,
even though with a large scatter. This is proved by the fact that
Model B -- where the two luminosities are completely unrelated -- is
rejected by the statistical tests.  In particular, the successful model
accounts for the dependence of the observed fractions of radio
detected quasars on apparent and absolute optical magnitudes, as due to
selection effects: going to optically fainter magnitudes (at a
fixed radio flux limit) corresponds to selecting increasingly more RL 
objects.
According to the model the intrinsic fraction of RL is small ($\simlt
5$~per cent) when compared to the total population, and this e.g. explains
the small fraction of radio detected quasars observed in the 2dF
sample. Since the shape of the $R^*_{1.4}$ distribution (see Figure
\ref{shape_R_P}) is steep for $1 \simlt R^*_{1.4} \simlt 10$, surveys
with brighter optical limiting magnitudes which select smaller values of
$R^*_{1.4}$ (at a given radio flux) will then result in more and
more radio detections. Similarly, the radio-optical dependence
accounts for the flattening of the faint end of the optical luminosity
function of RL quasars with respect to that of the whole population
(La Franca 1994; Padovani 1993).

Given the uniqueness of the solution, the main result of this work is
indeed the fact that we can put rather tight constraints on the
intrinsic radio properties of quasars.  The distributions shown in
Figure \ref{shape_R_P} could then describe the unbiased view of the
properties of the whole quasar population and this might possibly help
us to understand the physical mechanism(s) responsible for the radio
emission.  First of all, in the $R^*_{1.4}$ distribution we note no
lack or deficit of sources between the RL and RQ regimes: the
distribution has a peak at $R^*_{1.4} \sim 0.3$ and decreases
monotonically with a small fraction ($\simlt 5$~per cent) of objects
which are into the RL regime and represent a long tail of the total
distribution. This result contrasts (see also paper I) the view of a
RL/RQ dichotomy where a gap separates the two populations.
Nevertheless we can still talk about a ``dichotomy'' in the sense that
the data are compatible with an asymmetric distribution, with a steep
transition region and with only a small fraction of sources having
high values of $R^*_{1.4}$.

Then the basic questions are still open: do all sources belong to the
same population?  Or better: is there a single mechanism producing the
radio emission in quasars or two different processes dominate in the
bulk of the population and in the high $R^*_{1.4}$ tail?  While it is
believed that the radio emission in powerful radio quasars is produced
in well collimated jets (related to accretion processes), it is not
clear what is its origin in radio weaker and RQ sources. Let us then
consider some of the possible interpretations. \\

First of all it is known that, because of relativistic beaming, the
orientation of the source plays a leading role at least in the RL
regime -- although this could not account for the lack of large scale
radio structures in RQ sources.  In fact, relativistic boosting could
push up by typically a factor of $10^3-10^4$ the observed radio
emission (and correspondingly $R^*_{1.4}$ if the optical emission is
dominated by thermal radiation).  In this case we would expect at
least the extreme RL sources to be flat spectrum, which is not
supported by observations.  A large fraction of the RL sources in the
PG sample have a steep spectrum, and the same behaviour has been
suggested for RL objects from the FIRST-2dF sample (paper I).
Also well studied datasets of radio selected quasars,  from the 2-Jy 
(Wall \& Peacock 1985) and the 1-Jy samples (Stickel et al. 1994), 
show the distribution of radio-to-optical ratios to be the same for flat and 
steep spectrum radio quasars.
However, this result could be affected by variability because of the
non simultaneous radio flux measurements at different frequencies.

It has been proposed that the radio emission in RQ quasars is supplied
by ``starburst'' phenomena, i.e. thermal emission from supernova
remnants in a very dense environment (Terlevich et
al. 1992). Alternatively, the radio flux could be associated to
non-thermal emission from jets/outflows. In both cases we could reduce
the presence of a dichotomy to the capability of the central engine to
create a powerful/collimated jet. This in turn translates into the
quest for identifying the parameter(s)/physical condition(s)
responsible for its presence in only a few per cent of the sources.

As already mentioned, recent studies suggest the mass of the central
BH not to be tightly related to the radio emission (paper I; Woo \&
Urry 2002), except for a possible threshold effect (Dunlop et al.
2002) which however has been recently questioned (Woo \& Urry 2002).
Also, radio loudness does not seem related to the properties of
the host galaxy and of the environment (Dunlop et al.  2002), even
though the latter claim is still debated.  A further key physical
parameter for the nuclear activity is the mass accretion rate: however
it appears reasonable to assume that this is close to the Eddington
limit for all powerful optical quasars. Still open possibilities 
instead include the hypothesis that the creation of a jet is related
to a certain threshold in the BH spin or to the intensity and
configuration of the magnetic field in the nuclear regions.

A clue on the origin of radio emission comes from the radio imaging at
high resolution (pc scales) of RQ quasars, which revealed the presence
of non--thermal emission from jet-like structures, even for objects with low 
($R^*_{1.4} \sim 1$) radio loudness (Kukula et al. 1998; 
Blundell \& Beasley 1998). In many objects have been resolved
double or triple radio structures on scales of a few
kiloparsecs. Moreover, the inferred high brightess temperatures ($ T_B> 10^6 K$)
suggest that the radio emission cannot be produced by a starburst.
Then jets could
be a common feature in quasars and the radio power level could be due
to distributions in some of the jet properties, which however have to
present a `threshold' effect to reproduce the fast RQ/RL transition
and RL tail.  These properties could of course include the jet power
or  the efficiency of conversion of the jet energy into
radiation. It has been suggested (Rawlings \& Saunders 1991) that in
powerful radio sources the power released into the jet is comparable
to the accretion one. This would imply a radiative efficiency (in the
radio band) of only about 0.1 - 1 per cent for RL quasars -- and even
less in RQ ones (Elvis et al.  1994).  As a final remark we notice
that also time dependence/evolution could play an important role:
different ignition and duration times for the mechanisms responsible
for the optical and radio emission could explain the observed
distribution of radio-to-optical ratios.  We are currently
quantitatively comparing our results against these possible scenarios
and the findings will be the subject of a forthcoming paper.
%
\section*{Acknowledgments}
We are very grateful to Stefano Cristiani and Gianni Zamorani for
helpful discussions and we thank the referee for useful comments.
We acknowledge the Italian MIUR and ASI for financial support.

\end{document}